**JAMES CLERK MAXWELL'S CLASS OF 1856/57**

*John S Reid\**

*\*Department of Physics, Meston Building, University of Aberdeen AB9 2UE, Scotland. j.s.reid@abdn.ac.uk*

ABSTRACT

James Clerk Maxwell is known for his outstanding contributions to fundamental physics. These include providing the equations that govern electric and magnetic fields, establishing the basis of modern colourimetry, finding important relationships in thermodynamics, molecular science, mechanics, optics and astronomy. In his first Professorial chair in 1856 at the Marischal College and University of Aberdeen he undertook a substantial amount of teaching that laid the foundation for his later pedagogic output. This paper examines whom he taught, where his first students came from and what they did in later life, drawing material from a privately published memoir. Thumbnail portraits are included for 70% of his class. The analysis complements the usual emphasis on educational method and content. The data provide an interesting sociological survey of what Scottish University education was achieving in the middle of the 19th century and is presented as raw material for a wider enquiry.

INTRODUCTION

The passage of 150 years since James Clerk Maxwell was particularly productive has only enhanced his reputation as an outstandingly perceptive physicist. Comparatively recent articles and books aimed at both academics and a wider public continue to reveal aspects of his life, his contemporary influence and his legacy[1,2,3,4,5]. In 1856, at the age of 24, he was chosen as Professor of Natural Philosophy at Marischal College, Aberdeen[6] (more formally known as 'Marischal College and University'). The College was founded in 1593, the younger of two Universities in Aberdeen. The illustration here, reproduced from reference[7], shows it as Maxwell's class of 1856-57 saw it. Undoubtedly his significant academic achievements until then swung the University Court in his favour but in fact he was appointed to the University for his rôle as a teacher. During the academic year, Maxwell threw himself into this work. In his technical writing he acquired a reputation for great clarity of thought and for answering significant questions in his disciplines. It was totally in character that he tried to do the same in his teaching. The undergraduate textbooks he later authored were appreciated well into the twentieth century[8].

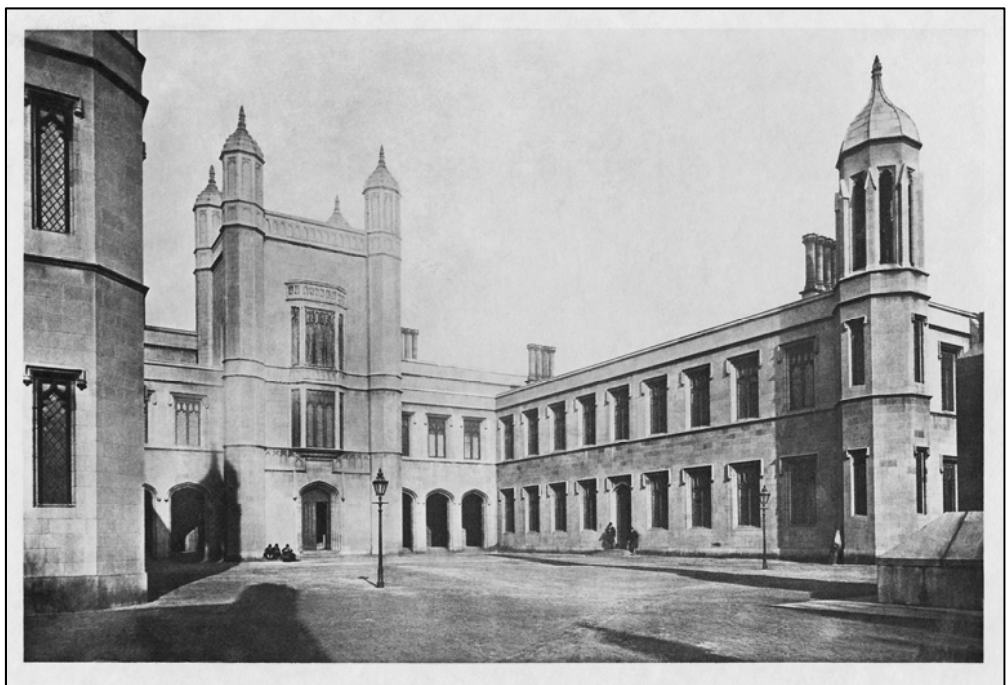

Nonetheless, his teaching success or otherwise is not the subject of this article. The question here is: who was all this effort aimed at?

Even in today's age of highly specialised University degrees, many graduates find career options in areas different from their specialism. The traditional Scottish University degree in Maxwell's day aimed to provide education rather than training, through a syllabus that included a broad mix of subjects ranging from the abstract through the worthy to the practical. Specialism was not its aim, yet the individual subjects were often taught by professors whose reputation rested on their advanced academic contributions.

We are fortunate in being able to find out who Maxwell taught in his first year as Professor, both from College records and from memoranda privately published for his class. The 1897 "Records of the Arts Class"[7] is a cumulative update of earlier versions produced in 1869 and 1879[9] by members of his class, with the addition of further career details, photographs of both class and College, and obituary reports. An analysis of these records makes a case study that complements the usual emphasis on educational method and content, and on the academic achievements of the professoriate. The information provides interesting social statistics on who attended University in the 1850s and what they did after their degree. It also illustrates the wide range of professionals who acquired an education in physics[10].

*THE CONTEXT*

Maxwell's course in Natural Philosophy occupied the greater part of a student's third year in a four year AM course at Marischal College. At least this was true for students aiming to come out with the Artium Magister degree (the modern Master of Arts) but the College also admitted 'private' students who could attend the courses of their choice without earning the right to graduate. In Maxwell's 1856-57 class there were 10 private students and 39 'gowned' students, as the regular students were called. Maxwell's first class began in early November 1856 and ended in early April 1857 with only a short break at New Year.

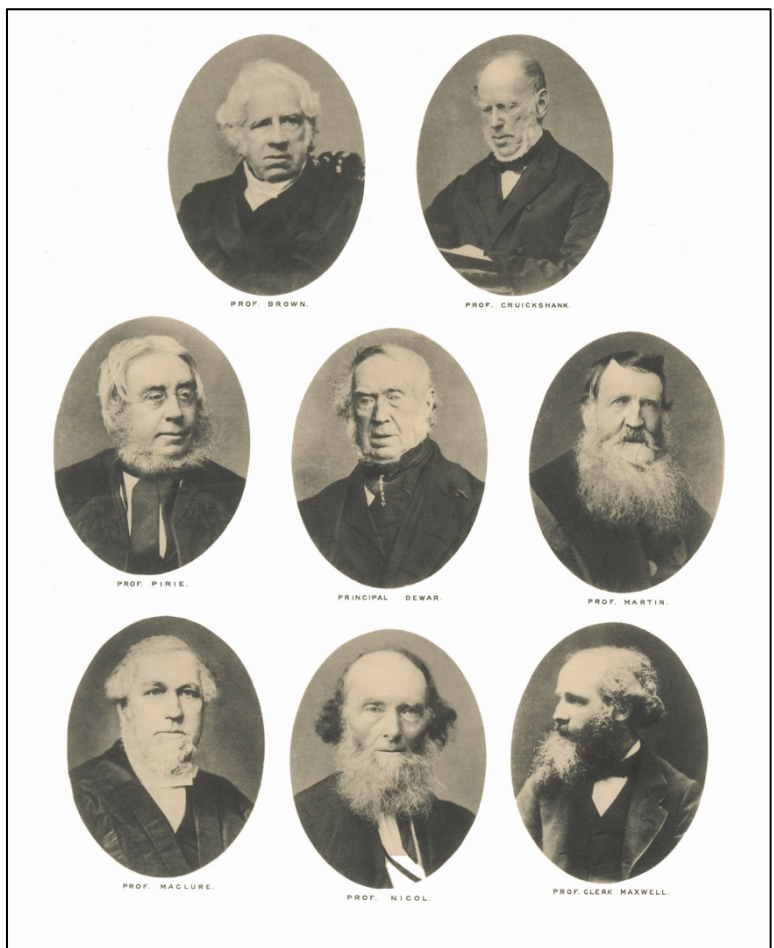

The adjacent Figure 2 shows the Professors who taught the compulsory courses for intending graduates during the 4-year degree. James Nicol (1810-1879), a perceptive geologist, was the only other scientist besides Maxwell. The senior professors in authority were called Regents, being Robert Brown (1792-1872), William Martin (1816-1890), James Nicol (1810-1879) and James Clerk Maxwell (1831-1879), each taking the principal class in one year of the course. The portraits are reproduced from reference[7].

There were optional classes too, some in other subjects taught by other staff but these did not count towards graduation. In his first year, Maxwell himself initiated an optional 4th year class in Natural Philosophy. Table 1 (alongside) shows where Maxwell's course fitted into the core degree programme taken by a gowned student. All years also attended Professor Pirie for one hour per week. The background of the professors involved has been summarised by Reid[11] and is not repeated here, for the focus of this article is on the students whom Maxwell taught in his main class.

*SOME STUDENT STATISTICS*

Table 2 on the next page shows the student names, their ages at the beginning of Maxwell's course, their places of birth ('Abdns'

| Year | subject | Professor |
|---|---|---|
| 1 | Humanity | Maclure |
| 1 | Greek | Brown |
| 2 | Humanity | Maclure |
| 2 | Greek | Brown |
| 2 | Natural History | Nicol |
| 2 | Mathematics | Cruickshank |
| 3 | Natural Philosophy | Clerk Maxwell |
| 3 | Mathematics | Cruickshank |
| 4 | Moral Philosophy and Logic | Martin |
| 4 | Evidences of Christianity | Dewar |
| all | Practical Religion | Pirie |

signifies Aberdeenshire) and the occupations of their fathers. 'G' flags the gowned students, 'P' the private students. Many dates of birth were given in the final version of the class record[7]. Some missing dates of birth were found on the 'Scotlandspeople' website[12] that gives public access to recorded pre-1855 births and baptisms. The first version of the class record[9] gave approximate ages 'as recorded in the Matriculation Books or Class Lists', a practice that seems to have begun in 1855. From this source the otherwise missing ages have been estimated for two gowned students and one private student.

| Category | Forename | Surname | Age (years) | Birth place | Father's occupation |
|---|---|---|---|---|---|
| G | James | Burr | 18.9 | Fyvie (Abdns) | Farmer |
| G | Patrick | Chalmers | 17.3 | Fyvie (Abdns) | Advocate |
| G | Alexander | Cochran | 16.4 | Aberdeen | Advocate |
| G | Charles | Cooper | 25.9 | Kincardine O' Neil (Abdns) | Farmer |
| G | John | Crombie | 17.7 | Newmacher (Abdns) | Textile manufacturer |
| G | William | Duguid | 15.4 | Udny (Abdns) | Farmer |
| G | James | Duncan | 17.9 | Arbroath (Angus) | Merchant |
| G | John | Duncan | 17.5 | Aberdeen | Advocate |
| G | William | Ewan | 19.0 | Aberdeen | Carpenter |
| G | Alexander | Fowlie | 17.1 | Strichen (Abdns) | Farmer |
| G | Angus | Fraser | 17.9 | Aberdeen | Merchant |
| G | Alexander | Frater | 16.2 | Aberdeen | City Chamberlain |
| G | Farquharson | Garden | 17.7 | Alford (Abdns) | Surgeon |
| G | Alexander | Gaull | 19.1 | Turriff (Abdns) | Merchant |
| G | William | Gordon | 17.6 | Abedeen | Stockbroker |
| G | John | Gray | 23.5 | Aberdeen | Weaver |
| G | William | Gray | 18.1 | Aberdeen | Merchant |
| G | Isaac | Henderson | 17.6 | Aberdeen | Architect/Builder |
| G | John | Hunter | 16.5 | Logie Buchan (Abdns) | Agricultural supplier |
| G | Alfred | Hutchison | 17.5 | Peterhead (Abdns) | Merchant |
| G | George | Mackenzie | 18.4 | Skene (Abdns) | Minister |
| G | James | Mackie | 24.2 | Laurencekirk (Kincardine) | Farmer |
| G | John | Macpherson | 17.6 | Tomintoul (Moray) | Minister |
| G | Joseph | Milne | 17.8 | Fetteresso (Kincardine) | Farmer |
| G | George | Morice | 17.9 | Edinburgh | Physician |
| G | George | Phillips | 18.5 | Aberdeen | Commercial traveller |
| G | George | Pickthorn | 17.4 | Aberdeen | RN Captain |
| G | John | Robertson | 16.7 | Aberdeen | Ironmonger |
| G | Robert | Simpson | 16.7 | Kintore (Abdns) | Minister |
| G | Simon | Simpson | 17.1 | Aberdeen | Advocate |
| G | James | Skinner | 19.1 | Rothiemay (Moray) | Mason |
| G | James | Smith | 18.4 | Auchindoir (Abdns) | Farmer |
| G | John | Smith | 17.1 | Aberdeen | Draper |
| G | William | Stables | 16.4 | Marnoch (Banff) | Vintner |
| G | James | Stuart | 20.7 | Rothiemay (Moray) | Farmer |
| G | Alexander | Thomson | 20.1 | Inverkeithny (Banff) | Farmer |
| G | Robert | Walker | 20.6 | Alford (Abdns) | Surgeon |
| G | James | Webster | 17.4 | Strichen (Abdns) | Innkeeper |
| G | William | Wilson | 17.2 | Ayr | Merchant |
| P | Stephen | Anderson | 13.7 | Elgin (Moray) | Shoemaker |
| P | Leslie | Clark | 25.2 | Aberdeen | Merchant |
| P | Alexander | Copland | 15.7 | Strichen (Abdns) | Merchant |
| P | Joseph | Hunter | 17.4 | Banchory Devenick (Kincardine) | Farmer |
| P | William | Jazdowski | 16.8 | Dungannon (N Ireland) | Professor of Languages |
| P | Richard | Lawrence | 15.9 | London | Commission Agent |
| P | William | Rattray | 15.5 | Aberdeen | Chemical works |
| P | Hector | Smith | 19.7 | Olrig (Caithness) | Landowner |
| P | James | Stewart | 35.8 | Ayr | Weaver |
| P | Robert | Sweeny | 16.5 | Montreal | Advocate |

There may be minor errors in some of the data, through ignorance of the students themselves or typographical slip, but the overall picture given by the statistics will be accurate enough.  The average student age at the beginning of Maxwell's course of the 39 gowned students is a little over 18 years[13].  Well over half of the class is clustered in the age range 16.2 to 19.1 years' old.  Today's third year classes at Aberdeen show a tight clustering but are some two years older.  The equivalent private students of Maxwell show a wider range of ages, from 13.7 years to 35.8 years and an average age that would be slightly younger but is skewed upwards by the one mature student.

Classing the father's occupations as either 'farmer', 'profession' or 'trade', 11 students (22%) were sons of farmers, 16 (33%) sons of professionals and 22 (45%) sons of tradesmen.  As shown later, almost all the students in Maxwell's class could be described as entering the professions after leaving University.

The majority of students were born in Aberdeen, the outlying shire or the adjacent counties of Kincardineshire, Banffshire and Morayshire (old boundaries).  Most would have grown up in the neighbourhood of their birth.  Sixteen (33%) hailed from Aberdeen and an equal number from Aberdeenshire.  Nine (18%) came from the adjacent counties; the remaining eight (16%) from further afield.  The prime function of the University was therefore to provide further education for the North-East of Scotland.  In this they were not alone in that King's College, Aberdeen, drew from the same catchment but its intake had a different flavour[14].  It had a slightly smaller enrolment.

Although all the gowned students were educated in ancient Greek and Latin, and of course read English, it is likely that the default social language of the class would have been conducted in the strong dialects of the North East of Scotland, a version of Scots still known as 'the doric'.  Maxwell's Scots was spoken with the different and softer Galloway accent but he would have had no difficulty in being understood in his informal contact with the students, such as at the social breakfasts he hosted.

*THE CLASS OF 1856/57*

The images in the next section put faces to some 70% of the class, namely 28 of the 39 gowned students and 7 of the 10 private students.  All the images except one come from the 1897 edition of the class records[7] and show the members as established adults, helping to personalise the statistics.  They are also a reminder that university education is not really aimed at the betterment of students but at the betterment of adults.  As might be expected, missing images are predominantly for those who had died by the time the 'Record'[7] was prepared.

Medicine, the law and divinity were classed as postgraduate subjects but it seemed in practice that it was not necessary for 4$^{th}$ year students to graduate to pursue one of these careers, or even for students to take the 4$^{th}$ year of the Arts course.  12 gowned students in Maxwell's class did not complete the 4 years of their study.  Following all the gowned students, Table 3 (below) illustrates that 7 qualified in medicine, 5 in law and 11 in divinity (one later on), a total of over half the students.  The 'divines' were spread over four denominations of the Protestant religion.   Seven of the divines went abroad, most of the lawyers and teachers practised locally.  Four students became schoolmasters, mainly teaching locally and four entered the Indian Civil Service.  In total 12 students made their careers locally, another five in Scotland and 20, half the class, spent much of their careers abroad.  India (6) was the most popular destination, though clearly a health hazard, followed by New Zealand (4).  Tables 2 (above) and 3 (below) together display more detail.  For example, sons of advocates mainly became lawyers; sons of farmers mainly became divines or doctors; sons of ministers of the church went abroad into the Indian Civil Service or the army.  The numbers are small but the correlations interesting.  24 (62%) of the class of gowned students are recorded as marrying.

*Table 3 (next page).*  Maxwell's students after leaving University.  G flags the gowned students, P the private students.  '**Location**' refers to the place where they pursued most of their professional careers, with 'Local' referring to Aberdeen and the surrounding counties.  '**Died**' refers to those who died before 1896.

The final version of the class records was drawn up in 1896, 40 years after Maxwell's class began their studies with him.  The last column of Table 3 shows those former class members who had died by then: 5 in the 1860s (generally men in their 20s), 6 in the 1870s (men in their 30s), 6 in the 1880s and so far 4 in the 1890s.  A 43% death rate over the 39 years since leaving the class shows that even professionals who could ostensibly better afford to look after themselves were seriously at risk in the last half of the 19$^{th}$ century.  Maxwell himself would reach only the age of 48.

| Category | Forename | Surname | Profession | Location | Died |
|---|---|---|---|---|---|
| G | James | Burr | Medicine | Local | 1893 |
| G | Patrick | Chalmers | Law | Local | 1889 |
| G | Alexander | Cochran | Law | Local | 1893 |
| G | Charles | Cooper | Divinity | India | |
| G | John | Crombie | Manufacturing | Local | |
| G | William | Duguid | Medicine | Local | |
| G | James | Duncan | Divinity | England | 1895 |
| G | John | Duncan | Law/Finance | Ceylon/New Zealand | |
| G | William | Ewan | Divinity | Local | |
| G | Alexander | Fowlie | Teaching | Local | |
| G | Angus | Fraser | Medicine | Local | |
| G | Alexander | Frater | Diplomacy | China | 1893 |
| G | Farquharson | Garden | Law | Local | |
| G | Alexander | Gaull | Divinity | Scotland/Abroad | ? |
| G | William | Gordon | Law | Local | |
| G | John | Gray | Divinity | Scotland/America | |
| G | William | Gray | Merchant | India | |
| G | Isaac | Henderson | Merchant | Burma | |
| G | John | Hunter | Teaching | England | |
| G | Alfred | Hutchison | Merchant | Hong Kong/China | 1876 |
| G | George | Mackenzie | HEICS/civil | India | |
| G | James | Mackie | Divinity | Scotland | |
| G | John | Macpherson | RE (army) | Abroad/England | |
| G | Joseph | Milne | Divinity | Scotland | 1871 |
| G | George | Morice | Divinity | New Zealand | 1884 |
| G | George | Phillips | | | 1860 |
| G | George | Pickthorn | Medicine/RN | d. New Zealand | 1866 |
| G | John | Robertson | HEICS/civil | India | 1873 |
| G | Robert | Simpson | HEICS/mil | India | |
| G | Simon | Simpson | RA(army) | Abroad/England | |
| G | James | Skinner | Teaching/Divinity | New Zealand | |
| G | James | Smith | Divinity | India | 1874 |
| G | John | Smith | Divinity | Natal | |
| G | William | Stables | Medicine/RN/author | South Africa/England | |
| G | James | Stuart | Medicine | Scotland | 1883 |
| G | Alexander | Thomson | Teaching | Local | 1862 |
| G | Robert | Walker | Medicine | England | |
| G | James | Webster | Teaching | Local | |
| G | William | Wilson | HEICS/civil | India | |
| P | Stephen | Anderson | Printer | USA | |
| P | Leslie | Clark | Civil Engineer | India | 1883 |
| P | Alexander | Copland | Teacher | Local | |
| P | Joseph | Hunter | Surveyor | Canada | |
| P | William | Jazdowski | Civil Engineer | Canada | 1885 |
| P | Richard | Lawrence | Accountant | England | 1882 |
| P | William | Rattray | Papermaking | India | 1865 |
| P | Hector | Smith | Farmer | New Zealand | 1878 |
| P | James | Stewart | Missionary | Local | 1874 |
| P | Robert | Sweeny | HEICS/mil | India | 1865 |

*GOWNED STUDENTS*

The following section gives brief details of the individual careers, sufficient to paint a broad picture. Information about marriage has been included to assist any genealogical interest. The entries below are largely a condensation of those in reference[7], though some typographical and minor errors in its text have been corrected and brief comments inserted in brackets.

| | | |
|---|---|---|
| 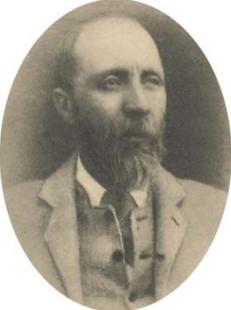 | *James Burr* | Omitted the 4th year. Studied medicine in Aberdeen and Glasgow, LFPSG, 1863; Edinburgh LRCP. Practised in his native parish of Fyvie (Aberdeenshire) before moving to Aberdeen later in the 1860s. Married Margaret Walker, shipmaster's daughter, 26th Sept. 1866. |
| 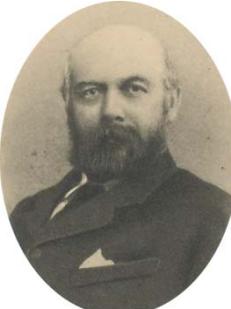 | *Patrick Chalmers* | Studied law in Aberdeen and Edinburgh. In 1859 joined Aberdeen Rifle Volunteer Corps in which he was a lieutenant. Joined the Society of Advocates in Aberdeen in 1864 and became a partner in the family legal firm of C. & J.H. Chalmers. This firm later became C. and P.H. Chalmers with Patrick as senior partner. Married Janie McDonell of Glengary in November 1880 and purchased the estate of Avochie (Aberdeenshire). |
| 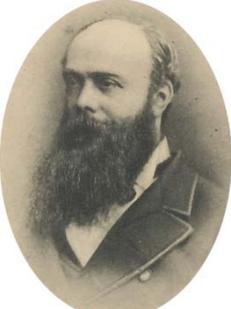 | *Alexander Cochran* | Studied law in Aberdeen and Edinburgh. In 1859 joined Aberdeen Rifle Volunteer Corps and rose to become colonel commanding the Volunteer Battalion Gordon Highlanders. Joined the Society of Advocates in Aberdeen in 1862 and became a partner in the family firm of Smith and Cochran, later becoming senior partner in its successors. Married Mary Campbell, daughter of Peter Campbell, Principal of the University of Aberdeen, on 24th Sept. 1867 and inherited the estate of Balfour (probably the Kincardineshire Balfour) from his father. |
| 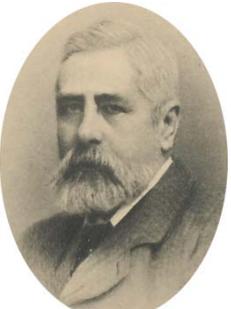 | *Charles Cooper* | Graduated AM 'with Honourable distinction' (the precursor of an Honours degree). Studied for 5 years in the United Presbyterian Divinity Hall in Edinburgh then acted as assistant to the Professor of Humanity in the University of Aberdeen. Ordained as U.P. minister at Holm of Balfron, Stirlingshire, in January 1866, a post he held for almost 3 years before being appointed head of Doveton Protestant College, Madras. In 1874 he was appointed Professor of Mental and Moral Science in the Christian College, Madras. |
| 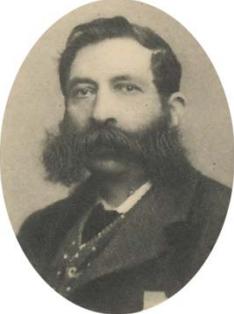 | *John Crombie* | Omitted the 4th year. Became a lieutenant in the Old Aberdeen Rifle Volunteers. Joined the family textile firm of J. & J. Crombie Ltd. (originator of the 'Crombie Coat'), Grandholm near Aberdeen and rose to become the senior partner. Married Annie Thompson of Pitmedden, daughter of George Thompson, sometime MP for the city of Aberdeen, on 22nd Aug. 1865. |
| 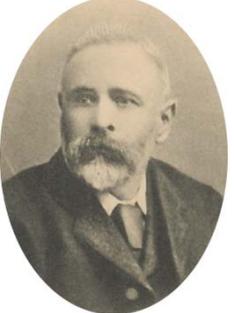 | *William Duguid* | Studied medicine in Aberdeen, obtaining MB (Batchelor of Medicine) and CM (Master of Surgery) in 1862 and MD (Doctor of Medicine) in 1875. Practised in Buckie (Aberdeenshire) where he became Lord Provost for six years. Medical Officer in the Volunteer Battalion Gordon Highlanders, becoming Surgeon Lieutenant-Colonel. Married Ellen Turner, farmer's daughter, 28th Dec. 1865. |
| *No image* | *James Duncan* | After Graduating AM 'with honourable distinction' spent four terms at Cuddesdon Theological College (now Ripon College), Oxford before being admitted to holy orders in the Church of England and appointed curate at Alverstock, Hampshire. In 1867 he translated to Christ's Church, Regent's Park, London and finally became Canon Residentary of Canterbury Cathedral. Married Cecile Palgrave, daughter of Francis Palgrave, Professor of Poetry at Oxford, in 1887. |
| 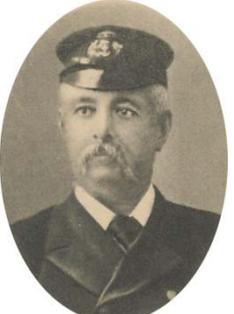 | *John Duncan* | Omitted the 4th year. Studied law and banking in Aberdeen before going to Ceylon in 1860 to Keir, Dundas & Co., merchants and coffee planters. In 1868 he established Duncan, Symons & Co., merchants in Colombo, Ceylon. In 1880 he emigrated to New Zealand where, after a while, he became a partner in the mercantile firm of Levin & Co., Wellington and for two years chairman of the Chamber of Commerce in Wellington. Married Emily MacCartney of Glenade, Ireland, on 28th April 1864. |
| 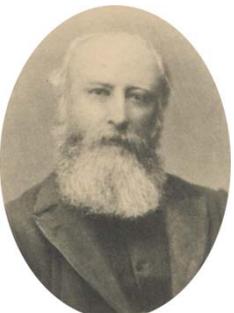 | *William Ewan* | Studied divinity at the Free Church College, Aberdeen, before becoming preacher of the gospel at the Free Church Presbytery, Paisley in 1862. After various appointments he became the Free Church minister at Fyvie (Aberdeenshire) in 1868 where he remained. Married Elisabeth Ramsay of Edinburgh on 15th Sept. 1869. |
| 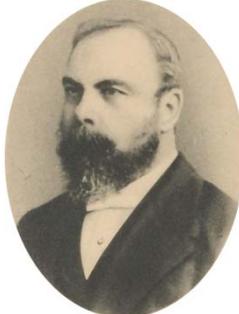 | *Alexander Fowlie* | In 1859 he became assistant teacher in Stewart's Hospital, Edinburgh and in 1861 parochial schoolmaster at Inverurie (Aberdeenshire). |
| 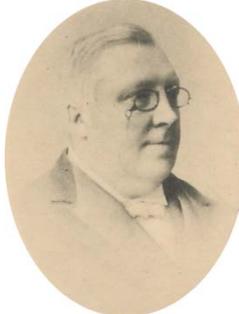 | *Angus Fraser* | Studied medicine in Aberdeen and Paris, becoming MD and CM of Aberdeen and FCS of London. Sometime assistant to Professor Brazier, University of Aberdeen Chemistry, and for many years practised in Aberdeen. Senior Physician and Lecturer on Clinical Medicine at the Aberdeen Royal Infirmary and for many years in the Aberdeen Rifle Volunteers, rising to Brigade Surgeon Lieutenant-Colonel in the Volunteer Battalion Gordon Highlanders. He represented the University of Aberdeen on the General Medical Council. (Angus Fraser's notes of Maxwell's lectures survive[15]). |
| 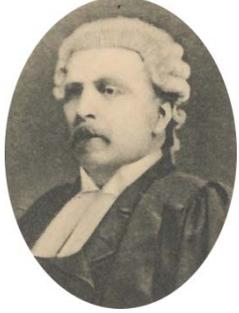 | *Alexander Frater* | Schoolmaster at Premnay (Aberdeenshire) before going to the War Office in 1861. From 1863 he had various appointments with the British Consular Service in China before being appointed Consul in Tamsui (Taiwan). In 1883, elected Barrister of the Middle Temple, London. In 1893 gazetted Consul in Hankow but had to abandon the posting due to ill health from which he died. Married Jessie Lawrie of Edinburgh in 1874. |
| 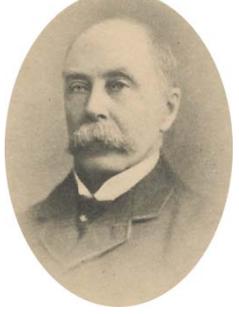 | *Farquharson Garden* | Omitted the 4th year. Studied law in Aberdeen and Edinburgh before being elected to the Society of Advocates in Aberdeen and becoming a partner in the law firm C. & J. H. Chalmers. For some time a lieutenant in the Administrative Battalion of Rifle Volunteers, Aberdeenshire. Married Mary Griffith, daughter of an insurance manager, in Aberdeen on 5th July 1870. |
| No image | *Alexander Gaull* | (**Gaul** on his birth certificate) studied divinity and became minister of the Church of Scotland at Hillside, near Montrose. He had left Hillside by 1877 and 'went abroad many years ago' and in 1896 was believed dead. |
| 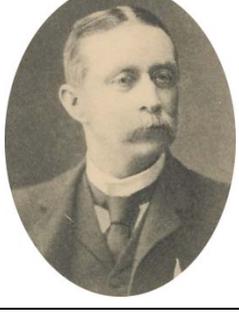 | *William Gordon* | Omitted the 4th year. Studied law in Aberdeen and appointed to the Society of Advocates in Aberdeen in 1864. Soon after he was appointed City Chamberlain of Aberdeen and in 1875 Town Clerk and Clerk to the Aberdeen Harbour Commissioners. Married Ella Paul of Waltham Cross on 21st June 1881. |
| 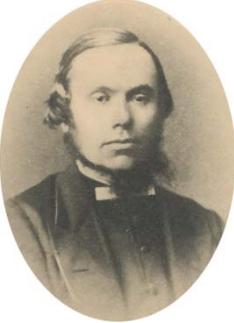 | ***John Gray*** | Studied divinity and became a minister of the Free Church of Scotland. He had assistantships at St Peter's Church, Dundee and then Hutcheson Free Church, Glasgow, before being ordained minister at Gorebridge near Edinburgh in 1866. He had left Gorebridge by 1877 and emigrated to America. |
| 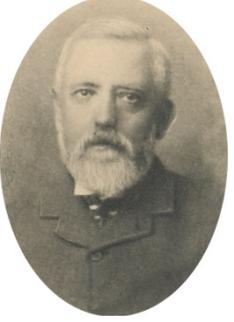 | ***William Gray*** | After graduating he worked with A. & W. Thom, merchants, Aberdeen, before going to the Peninsular and Oriental Co., London (now P&O). In 1864 he went to Hong Kong in service of the company and then to Bombay. In 1875 he retired from the company and returned to England and then to Scotland. In 1869 he married Louisa Butt, daughter of a London merchant. |
| 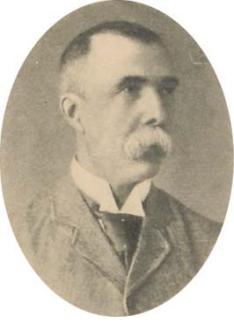 | ***Isaac Henderson*** | After graduating he entered a merchant's office in Glasgow where he stayed until 1864 before emigrating to Moulmein in Burma (now Mawlamyine in Myanmar). In 1868 he co-founded Hannay, Henderson & Co. there. On being joined by his brother this became Henderson & Co. and in 1896 was still active. |
| 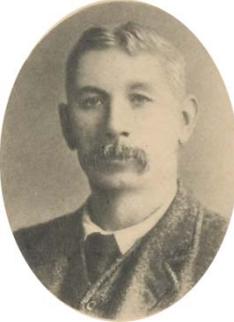 | ***John Hunter*** | After graduating he became assistant master, first in Blackburn, Lancashire, then Salway House, Leyton and afterwards in Spring Grove School, London. He spent 9 years at Holly Bank School, Cheetham Hill, Manchester before becoming a master at Ewart High School, Newton Stewart. |
| *No image* | ***Alfred Hutchison*** | Omitted the 4$^{th}$ year. In 1860 went to China with Turner & Co., tea merchants, Hong Kong and in 1871 became a partner in Deacon & Co., Canton. Died in London in 1876. Married Mary Jane Adams, daughter of an Indian Army MD, on 10$^{th}$ Dec. 1874. |
| 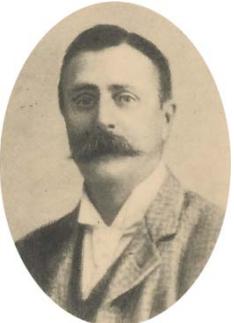 | ***George Mackenzie*** | Omitted the 4$^{th}$ year and in 1857 obtained a commission as Ensign in the HEICS (Honourable East India Company Service), Bombay, in the 2$^{nd}$ European Regiment, Light Infantry. In 1870 he joined the Staff Corps and held appointments at Hyderabad and Mysore in the Revenue, Survey and Assessment Department. He retired as Colonel after 26 years of service, to St. Heliers, Jersey. Married Mary Reid of Aberdeen in 1865. |
| 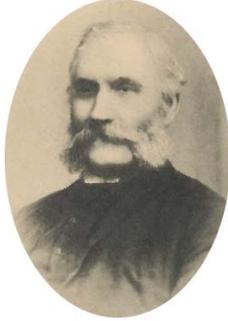 | ***James Mackie*** | Studied divinity becoming a minister in the Church of Scotland. He began as assistant in St George's, Edinburgh and then minister of St. Mary's, Partick, Glasgow, and then St Mary's, Dumfries. His career took him to at least half a dozen further postings, including the year of 1880 spent as assistant minister of St Paul's, Montreal. He married Henrietta Rollatt on 16$^{th}$ Dec. 1864 and after her death in 1882 he re-married on 20$^{th}$ Nov. 1894 Marian Morton of Annan who died some 10 months later. |
| 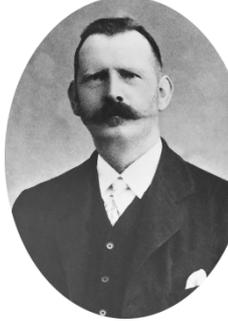 | ***John Macpherson*** | Obtained a commission as 2$^{nd}$ lieutenant in the Royal Engineers in 1859 and was stationed for some time at the Cape of Good Hope. On his return, appointed to the Ordnance Survey in 1872 and from 1875 to 1878 surveyed the Hebrides and Shetland. For the three following years he was posted to Bermuda. Upon his return he rose to become Director General of the Survey. In 1890 he was created CB, in 1893 promoted to Colonel in the Royal Engineers (and knighted KCB in 1899. He inherited from his uncle on his mother's side of the family the Corrachree estate near Tarland (Aberdeenshire) in 1888 and then assumed as surname her maiden name of Farquharson. Hence as Director General from 1894 – 1899 he was Brevet Colonel John Farquharson). Image © Crown copyright Ordnance Survey. |
| 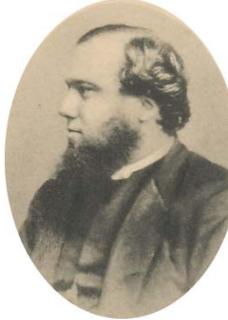 | ***Joseph Milne*** | Graduated 'with honourable distinction', studied divinity becoming a minister of the Church of Scotland. Acted as assistant in the Cathedral, Glasgow before becoming minister in the parish of Bathgate in 1869. Died in 1871, aged 32. Married Mary Burns, daughter of the Minister of Auchtergaven, on 26$^{th}$ April 1870. |
| *No image* | ***George Morice*** | Omitted the 4$^{th}$ year. Studied divinity before becoming a minister of the Free Church of Scotland. After a time as assistant in Kintore (Aberdeenshire) in 1866 he sailed to New Zealand and was appointed to the Presbyterian Church in Napier where he remained for five years. Returned to Scotland to marry and went back to New Zealand to posts in Hokitika, Christchurch and finally Otago. He was drowned in 1884 attempting to rescue his wife. Married Williamina Barclay of Auldearn (Nairnshire) in 1873. |
| *No image* | ***George Phillips*** | Died in Glasgow in 1860. |
| 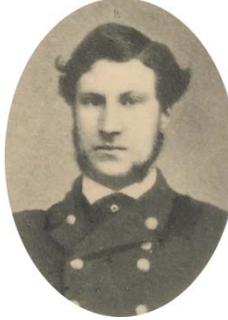 | ***George Pickthorn*** | Studied medicine, obtaining his MB in Aberdeen in 1862. Followed his father's career by joining the Royal Navy, as assistant surgeon aboard SS "Trafalgar" in the Mediterranean and SS "Achilles" in the Channel before being appointed to HMS "Challenger" in the Australian Squadron. (Died aged 27 at Auckland, New Zealand and is buried in Symonds Street cemetery. His name is recorded on a commemorative brass plaque). |
| 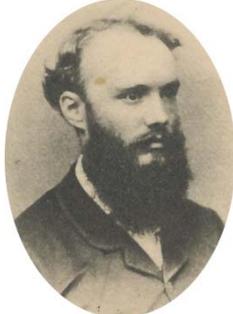 | *John Robertson* | Graduated 'with honourable distinction'. Entered the Indian Civil Service in 1861 and appointed Assistant Magistrate and Collector in Futtehgurh. Following promotions he became Joint Magistrate and Collector at Moradabad in 1873 but died in Futtehgurh that year, aged 33. Married Isabella Grant, daughter of the minister of the Small Isles, 25th Feb. 1867. |
| *No image* | *Robert Simpson* | Omitted the 4th year. Entered the HEICS (Madras Presidency, 36th Native Infantry) in Feb. 1858. With some breaks home for ill health he rose through the ranks to Captain, Major, Lieutenant-Colonel and finally in 1888 Colonel when he retired. Married Elisabeth Bisset, daughter of the minister of Bourtie (Aberdeenshire), on 7th June 1888. |
| 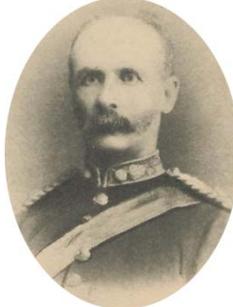 | *Simon Simpson* | Omitted the 4th year. Entered the Royal Artillery where he served in Canada, Malta, India and the West Indies. Promoted to Captain in 1878, Major in 1879, serving in Gibraltar and Bermuda. As Lieutenant-Colonel in 1886 he served at Singapore and Straits Settlements before commanding the Royal Artillery Depot at Seaforth near Liverpool. He retired as Colonel in 1891. Married Julia Deare, daughter of Colonel Deare, on 10th June 1875. |
| *No image* | *James Skinner* | Began as a schoolmaster in England and then in Elgin (Morayshire) but gave up teaching to enter a merchant's office in Manchester. He then studied divinity and became minister of the Free Church Presbytery of Strathbogie (Aberdeenshire) before emigrating to New Zealand where he was minister of the Presbyterian Church in Havelock and finally Flemington (S. Island). |
| *No image* | *James Smith* | Studied divinity and became a minister of the Church of Scotland obtaining his first charge at Alford (Aberdeenshire) in 1863. Soon after he sent as a missionary to Madras but had to return in poor health. He was admitted to the Royal Lunatic Asylum in Aberdeen in 1865 and died there in January 1874. |
| *No image* | *John Smith* | Graduated 'with honourable distinction' and studied divinity becoming a licentiate of the Free Church of Scotland in 1862. After assistantships in Fraserburgh (Aberdeenshire) and Fountainbridge, Edinburgh, he went to South Africa in 1865 and was ordained by the Presbytery of Natal. In 1870 he became the minister at St John's, Pietermaritzburg where he was also the first president of the Pietermaritzburg YMCA. |
| 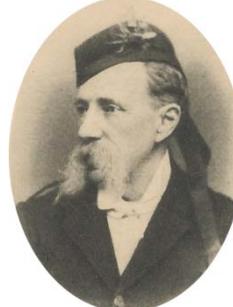 | *William Stables* | Omitted the 4th year. Studied medicine in Aberdeen obtaining his MD and CM in 1862. He entered the Royal Navy in 1863 as Assistant Surgeon and was appointed to HMS "Narcissus" on the Cape of Good Hope and African Station. Invalided out in 1875 he settled in Twyford and in the 1880s began a literary career writing stories and novels of adventure and travel (under the name Gordon Stables. His book "From Ploughshare to Pulpit" is based on his Marischal College days and includes passing accounts of his Professors, Maxwell included). In 1864 he married Theresa McCormack, daughter of Captain McCormack of Pembrokeshire. |
| *No image* | *James Stuart* | Studied medicine and obtained his MD and CM in 1863. He practised at New Pitsligo (Aberdeenshire) and in Newcastle-upon-Tyne before taking up a post at Bucklyvie in Stirlingshire where he was Medical Officer for Health. He then moved to Canisbay, Wick. He married Margaret Fairburn, daughter of an iron merchant, Newcastle-upon-Tyne, on 23rd Sept. 1868. |
| *No image* | ***Alexander Thomson*** | Became assistant master in Dr Tulloch's Academy, Aberdeen, and in 1860 parochial schoolmaster at Drumblade (Aberdeenshire). He died in 1862. |
| 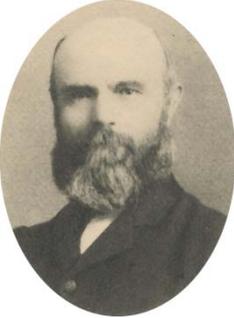 | ***Robert Walker*** | Studied medicine at Aberdeen, obtained his MD and CM in 1863 and established a practice in Wooler, Northumberland. He married Jane Atkinson of Ewart on 24th July 1866. |
| 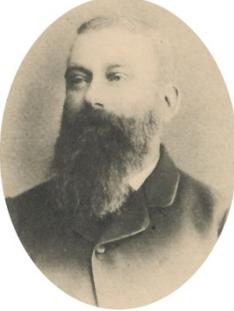 | ***James Webster*** | Studied divinity but took up teaching, first in Higher Broughton, Manchester, then in Kincardine O'Neil (Aberdeenshire) before moving to Shannas, Old Deer (Aberdeenshire). He married Christina Ross, farmer's daughter from Lumphanan (Aberdeenshire), on 9th Sept. 1880. |
| 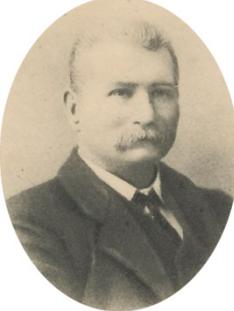 | ***William Wilson*** | Graduated 'with honourable distinction' and won a place in the Civil Service of India in 1859 and in the following year joined the Madras Presidency. He served in various districts before becoming Judge of the Kistna District in 1874. A series of posts from Collector and Magistrate of Colmbatoor through Member of the Board of Revenue, Director of Revenue Settlement led to him becoming Revenue Secretary to the Government and in 1885 Chief Secretary. He was appointed Fellow of the University of Madras. He married Alice Cooper, daughter of a Madras High Court judge, on 8th July 1869. |

There are many interesting asides in the detail. One of the class married the Principal's daughter, as Maxwell had done with the previous Principal. At the other end of the scale of fortune, one ended up in a lunatic asylum where he spent 9 years of his comparatively short life. One ended up as Professor of Mental and Moral Science, another, the author of some 130 books, mainly children's adventure stories[16].

*THE PRIVATE STUDENTS*

| | | |
|---|---|---|
| 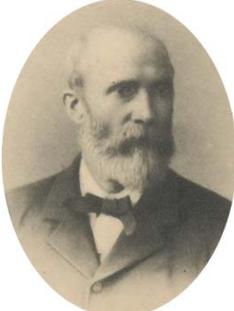 | ***Stephen Anderson*** | Learnt the trade of printing from William Bennett, Aberdeen, and emigrated to Baltimore, USA, to work for Innes & Co., printers. |
| 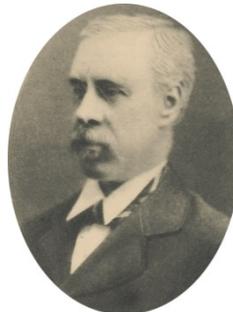 | *Leslie Clark* | Studied civil engineering with Messrs Abernethy, London, before emigrating to Ceylon in 1864 as an assistant engineer with the Ceylon Railways. He moved to India where, amongst other projects, he was district engineer with the Great Indian Peninsular Railway Co. engaged in the construction of the Wardha and Wadwan lines. In 1875 he designed and oversaw the construction of the Lahore waterworks and drainage system, the first of its kind in the Punjab. Married Sarah Matthews, London, Oct. 1862. |
| 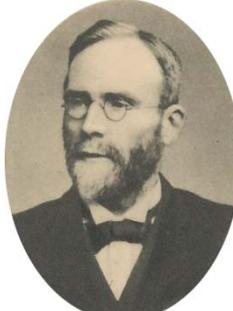 | *Alexander Copland* | In 1860 he began a teaching career in England as assistant master with a series of appointments in Holloway, Hastings and Blackheath. He then returned to Aberdeenshire to spend most of his career in Fraserburgh. He was elected President of the Deeside Field Club in 1894. Married Isabella Clark, Fraserburgh, 23$^{rd}$ Dec. 1869. |
| 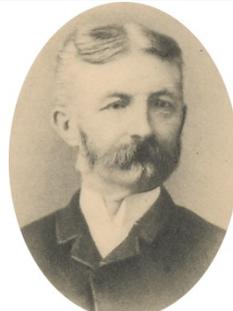 | *Joseph Hunter* | Studied land surveying with J. F. Beattie, Aberdeen, and in 1864 emigrated to British Columbia. He settled in the Cariboo region and was its representative in the first Provincial Parliament until 1875. In 1872 he joined the engineering staff of the Canadian Pacific Railway and rose to divisional manager. In 1884 he was appointed Chief Engineer for the Esquimalt and Nanaimo railway (Vancouver Island) whose route, construction and superintendence he oversaw. |
| No image | *William Jazdowski* | Studied civil engineering with A. Gibb, Aberdeen, then joined the North-Eastern Railway Co. before moving to the London & NW Railway Company. In 1872 he emigrated to Canada and became engaged on the first survey of the Canadian Pacific Railway. For several years he was head of the survey office in Winnipeg. Married Jessie Milne, daughter of the rector of Dollar Academy, in Aug. 1881. |
| 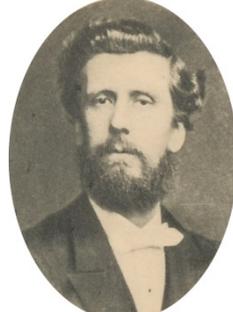 | *Richard Lawrence* | He returned to London as accountant for a London mercantile house. He became an active Baptist in 1862, author of religious tracts and pastor at the Strict Baptist Chapel, Bermondsey, London. Married Annie Lake, Bermondsey, on 24$^{th}$ Dec. 1864. |
| No image | *William Rattray* | Studied papermaking with Messrs Pirie, Stoneywood (Aberdeen) and went to Bombay to manage a paper mill. He had to return home in ill health and died aged 24. |
| 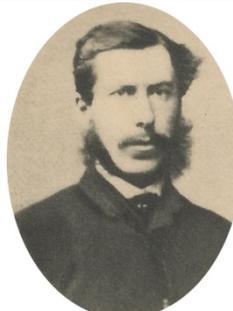 | *Hector Smith* | Emigrated in 1858 to New Zealand to become a sheep farmer. Returned briefly in 1866/67 when he married Anne Barron, Aberdeen, 22$^{nd}$ Jan. 1867, before going back to New Zealand. Died near Napier aged 41, having survived his wife by a few months. |
| 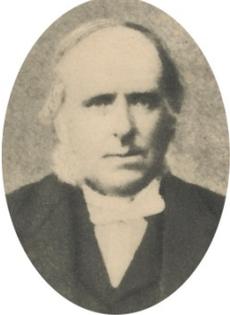 | *James Stewart* | Was a City Missionary with the fishing village of Footdee, Aberdeen, and the first missionary connected with the Sailors Institute, Aberdeen. He had married Elspeth White on 29$^{th}$ Dec. 1841. |
| No image | *Robert Sweeny* | Joined the HEICS (Bengal Presidency) and was for some time lieutenant in the Light Cavalry. Returned home in 1864 and married. Went back to India in 1864 but died the following year. |

Eight out of 10 private students had previously enrolled in the second year Mathematics course. 7 out of 10 took the third year Mathematics course in addition to Maxwell's course; 7 out of 10 went abroad; 7 out of 10 married; 7 out of 10 died without reaching the age of 55. James Stewart, the only married student, enrolled in private classes in all four years. As a whole these were a different set of young men from the gowned students, with a bias in careers towards the industrial world.

In Britain, only King's College, London, and Glasgow University had Professors of Engineering in the 1850s yet it is clear that the 'private student' arrangement in the Scottish Universities provided valuable education for would-be engineers, quite sufficient for them to pursue their interest. In addition, Maxwell also ran an evening class largely for those of an engineering or mechanical bent already employed in the city. This course has been discussed by Reid[11].

*CONCLUSION*

Maxwell was teaching at a time when Scottish scholarship had a distinctive profile. In two of the four years of the general degree course, science was the major subject – Natural History in year 2 and 'Natural Philosophy' in year 3. Future lawyers, ministers of the church, doctors, school-teachers, accountants and military and colonial administrators, to cite some of the careers of Maxwell's pupils, all received basic literacy in scientific knowledge and, equally importantly, in the methods of acquiring knowledge that are practised in science. It was a regime that Maxwell was very familiar with through his own student experience in Edinburgh. Judging by the effort he devoted to his teaching[6, 11], it was one he approved of. Times were changing, though, with the Universities (Scotland) Act 1858 spelling out revised Government and Discipline of the Universities of Scotland and leading the way to increased specialisation. It was this same Act that spelt out the union of the two universities in Aberdeen that would cost Maxwell his Chair. The careers of Maxwell's pupils are a record of what the old system achieved.

The cross-section of student careers found in this particular class is quite representative of those who attended Maxwell's Marischal College classes in later years[10]. The class that enrolled later in 1857 included George Reith, father of J C W Reith, the guiding light of the BBC from its foundation in 1922 until his resignation in 1938. George Reith became a minister of the Free Church of Scotland, as had five of Maxwell's 1856/57 class. In Maxwell's final year, among the gowned students was George Croom Robertson (who in 1866 became Professor of Mental Philosophy and Logic at University College, London) and, as a private scholar, Maxwell's only student who became a distinguished scientist. This was David Gill, perhaps Scotland's most successful astronomer. These two were exceptions. Maxwell himself would have been in no doubt that the aim of the College in general, and his course no less than others, was not to train future scientists but to produce educated adults. In this the College was certainly successful.

It is not entirely a stroke of good fortune that we can trace Maxwell's students, for their cohesion as a social group is also a reflection of the value the class as a whole placed on their experience at Marischal College. The sense of community engendered by their Alma Mater is one that many Universities today would be pleased to match. If the other Scottish Universities produced a comparable experience, we may be able to find how Maxwell's students compared with the roughly contemporary cohorts of James D. Forbes and his successor P. G. Tait in Edinburgh, with those of William Thomson in Glasgow or Thomas Jackson and his successor William Swan in St Andrews.

*Notes*

1       Ed. John S. Reid, Charles H.-T Wang and J. Michael T. Thomson "*James Clerk Maxwell 150 years on*" a special issue of Phil. Trans. Roy. Soc. A 366, pp 1649-1874 (2008) devoted to Maxwell's life and legacy.

2       Ed. David Forfar "*Celebrating the Achievements & Legacy of James Clerk Maxwell*" The Royal Society of Edinburgh (Edinburgh, 2008).

3       Raymond Flood, Mark McCartney and Andrew Whitaker "*James Clerk Maxwell; Perspectives on his Life and Work*", OUP (Oxford, 2014).

4       Basil Mahon and Nancy Forbes "*Faraday, Maxwell and the Electromagnetic Field: How two men revolutionised physics*" Prometheus Books (New York, 2014).

5       Robyn Arianrhod "*Einstein's Heroes: Imagining the World through the Language of Mathematics*" OUP (Oxford, 2013).

6       This and many other details of Maxwell's Marischal College teaching are discussed in John S. Reid "*Maxwell in Aberdeen*" pp 17 – 42 in reference[3].

7       Ed. Farquharson Taylor Garden "*Records of the Arts Class 1854-58, Marischal College and University*" (Aberdeen, 1897). This private publication drew on College records and added many personal details of the class.

8       James Clerk Maxwell authored three undergraduate level textbooks: "*Theory of Heat*" (Longmans, Green & Co., London, 1st edition, 1871; 18th edition ed. Lord Rayleigh, 1921 and later reprints); "*Matter and Motion*" (SPCK, London, 1st edition 1876; new edition ed. J. Larmor, 1920 and later reprints); "*An Elementary Treatise on Electricity*" (OUP, ed. W. Garnet, 1881, 2nd edition, 1888 and later reprints). All these books also appeared in translation.

9       Ed. PHC [Patrick Henderson Chalmers] "*Records of Bajeant Class: Marischal College Session 1854-55*" (1869) private publication, in Aberdeen University Library; 2nd edition revised by John Crombie, 1879.

10      No other 'record book' produced by graduates in Maxwell's other three years at Marischal College has come to light. However Peter John Anderson "*Fasti Academiae Mariscallanae Aberdonensis*" Vol II, (The Spalding Club, Aberdeen, 1898) includes very brief career notes for many of the students in all four years of Maxwell's teaching at Marischal College. These were provided by members of each class.

11      John S. Reid "*James Clerk Maxwell's Scottish Chair*" Phil. Trans. Roy. Soc. A (2008) 366, pp 1661-1684.

12      http://www.scotlandspeople.gov.uk/welcome.aspx, accessed 1st December 2014.

13      Reid in reference[11] gave the figure of 18 years and 9 months but this was based on the integer ages given in reference[9] and not the dates of birth used here.

14      The nearest King's College (Aberdeen) record is that of the Arts Class 1856-60 (privately printed, Aberdeen, 1906), contemporary with Maxwell's 3rd year of teaching. Although this publication is much longer than the Marischal Record[8] it is less complete in the detail of student origins. Of the 40 students who definitely took the Natural Philosophy Class, only one came from Aberdeen city, 18 from Aberdeenshire, 11 from adjacent counties and 10 from further afield. About 50% of the occupations of the fathers where recorded were farmers.

15      A. Fraser (1857), "*Lecture notes of Professor Maxwell taken by Angus Fraser on Natural Philosophy*", Aberdeen University Special Libraries and Archives Ms AMCS/3/17.

16      Wikipedia includes a brief biography at http://en.wikipedia.org/wiki/William_Gordon_Stables, accessed 11/12/2014. 107 of his books are listed at http://en.wikisource.org/wiki/Author:William_Gordon_Stables, accessed 15/12/2014.